\documentclass[preprint2]{aastex}
\usepackage{amsmath, amssymb}

\def\Vec#1{\mbox{\boldmath $#1$}}

\slugcomment{Published in {\it NAIS J.}, {\bf 7}, 5--16 (2012)}

\shorttitle{The Fast Invariant Imbedding Method Coupled With the Doubling--ADDING  Method }
\shortauthors{Kawabata}

\begin{document}


\title{A HYBRID ALGORITHM OF FAST INVARIANT IMBEDDING  AND DOUBLING--ADDING METHODS FOR EFFICIENT MULTIPLE SCATTERING CALCULATIONS}


\author{Kiyoshi Kawabata}
\affil{Department of Physics,  College of Science, Tokyo University of Science,  \\
    Shinjuku-ku, Tokyo 162-8601, \\ Japan}
\email{kawabata@rs.kagu.tus.ac.jp}
 \begin{abstract}
\thispagestyle{empty}
An efficient hybrid numerical method for multiple scattering calculations is proposed.  We use  the well established doubling--adding method to find the reflection function of the lowermost homogeneous slab comprising the atmosphere of  our interest. This reflection function provides  the initial value for the fast invariant imbedding method of  Sato {\it et al.}\,(1977), with which layers are added until  the final reflection function of the entire atmosphere is obtained.
The execution speed of this hybrid method is  no slower than one half of  that of the doubling-adding method, probably the fastest algorithm available,  even in the most unsuitable cases for the fast invariant imbedding method.
The efficiency of the  proposed method increases rapidly with  the number of  atmospheric slabs and the optical thickness of each  slab.  For some cases, its execution speed  is approximately four times  faster than the doubling--adding method.  \vspace{0.3cm}\\
\end{abstract}


\section{Introduction}
Accurate and sufficiently rapid multiple scattering  calculations of the intensity distributions of  solar radiation reflected by planetary atmospheres are essential for performing  the  terrestrial and planetary remote sensing studies. 
Many  methods for performing multiple scattering calculations have  been proposed (e.g., Hansen and Travis, 1974; Natsuyama {\it et al}. 1998;  Liou, 2002; Hovenier {\it et al.} 2004;  Mishchenko {\it et al}. 2006). The {\it invariant imbedding } method is one such method, and it derives a set of  integro-differential equations for reflection and transmission functions. It yields these equations by considering the change in the intensity of  outgoing radiation when a very thin slab  with given optical properties  is added either to the top or   bottom of the main body of the atmosphere. Although the equations thus derived are the same as those derived by Chandrasekhar (1960) by means of the invariance principle, the invariant imbedding method provides us with a short cut to arrive at them; therefore we call them {\it  invariant imbedding equations}. \par
The {\it doubling} method, in contrast to the invariant imbedding method, finds reflection and transmission functions for a stack of two identical homogeneous layers whose reflection and transmission functions are known.  First,  a slab of sufficiently small optical thickness is considered, so that its reflection and transmission functions can be well approximated by single and  second--order scattering solutions, which are simple.
By repeating this doubling procedure, the reflection and transmission functions of any homogeneous atmosphere of arbitrary optical thickness can be produced.  Note that  the doubling method is a special case of  the {\it adding} method, where the reflection and transmission functions of a stack of two slabs of different optical properties are sought.\par
One important problem associated with solving the invariant imbedding equations is that they belong to a class 
 of  so-called  {\it stiff differential equations}.  Therefore,  it is extremely difficult to numerically integrate them with any standard technique such as the Runge--Kutta method even with an extremely small step size.  \par
The {\it fast invariant imbedding} method of Sato {\it et al.} (1977) circumvents this problem by approximating the source term of each equation by low order polynomials of  optical height $\tau$ measured upward from the ground surface. This approximation is  initially a linear function followed by a piecewise quadratic polynomial of   $\tau$. 
These equations are then integrated semi-analytically over 
 $\tau$. As a result, we obtain a set of nonlinear 
implicit equations for  the reflection and transmission functions at each integration step.  These equations   can then be solved directly by  successive iterations.\par
However, the fast invariant imbedding method still tends to be several times slower 
than the doubling--adding method for atmospheres of moderate or large optical thickness.  In this study, we therefore 
 attempt to  improve the computational efficiency of the fast invariant imbedding method by incorporating the doubling--adding method to initialize the reflection and transmission functions of the lowermost layer.
\section{Formulations}
\subsection{Basic Equations}
For simplicity, we ignore the effect of polarization of radiation, so that the scalar approximation of the relevant quantities is valid. Let us also restrict our argument primarily to the computational aspect of the reflection function in view of  remote sensing applications. Furthermore, we assume that the entire atmosphere of optical thickness $\tau_{\rm T}$ is suitably approximated by $N$ homogeneous slabs, with the first slab being the lowermost, and the $N$-th slab being the topmost as in Kawabata and Hirata\,(1985). In addition, we assume 
 that the ground  acts like a Lambert surface of reflectivity $A_{\rm grd}$, which isotropically 
reflects  incident light. \par
Let us measure the optical height $\tau$ of a given location from the ground,
 because we intend  to build the atmospheres of  interest by stacking slabs upward. 
Furthermore, let  $\Delta\tau_n$ denote the optical thickness of the $n$-th slab. Then,  the total  optical height $\tau_{{\rm tot},n}$ of  the upper surface of the $n$-th slab is given by
\begin{equation}
\displaystyle{\tau_{{\rm tot},n}=\sum_{j=1}^n\Delta\tau_j}.  \label{eq-1}
\end{equation}
Hence, $\tau_{{\rm tot},N}=\tau_{\rm T}$, i.e., the total optical thickness of the entire atmosphere as shown in Fig. 1. \par
\begin{figure}[!htb]
\centering
\hspace*{0.6cm}\includegraphics[width=0.98\linewidth]{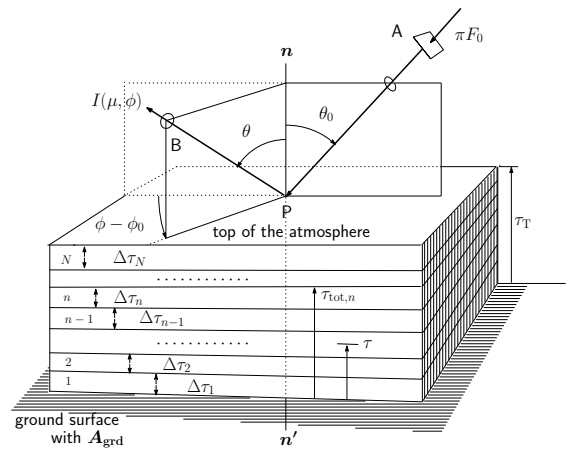}
\caption{Geometry for multiple scattering calculations.
 The incident light with  flux $\pi F_0$ per unit area\,(designated by A) perpendicular to the direction of incidence given by the zenith angle $\theta_0$ and the azimuth angle $\phi_0$ enters a point P on the top of the atmosphere, and emerges with the intensity $I(\mu,\phi)$ in the direction B specified by the zenith angle $\theta$ and the azimuth angle $\phi$.  
 The atmosphere is  plane-parallel and scattering--absorbing, and it has total optical thickness $\tau_{\rm T}$. The zenith angles  $\theta$ and $\theta_0$ are  measured from the local upward normal $\mbox{\bf n}$, and the azimuth angles $\phi$ and $\phi_0$ are measured counterclockwise when the upper surface of the atmosphere is observed from above.
The atmosphere is assumed to be stratified with  $N$ homogeneous slabs of optical thickness $\Delta\tau_n\ \ (n=1, 2, \cdots, N)$, and it is bounded at its bottom by a Lambert surface having  reflectivity $A_{\rm grd}$.  The straight line $\mbox{\bf n}^\prime$ indicates the downward normal perpendicular to the bottom surface.
 \label{fig1}}
 \end{figure}%
 The intensity $I(\mu, \phi)$ of radiation reflected from a plane-parallel atmosphere can be
specified by zenith and azimuth angles. The zenith angles  $\theta$ and $\theta_0$ measure 
the radiation  emergence  and incidence  directions, respectively, with respect to the outward normal $\Vec{n}$ to the upper surface of the atmosphere in question. The azimuth angles $\phi$ and  $\phi_0$ for these two directions, respectively, are measured counterclockwise when the upper surface 
 is observed from above.\par
For a mono-directional light incident from a direction  $(\mu_0, \phi_0)$, 
 the intensity of reflected light $I(\mu, \phi)$ emerging from the atmosphere in the direction $(\mu, \phi)$ measured with respect to the upward normal $\Vec{n}$ to the topmost surface (Fig. 1) can be  expressed in terms of the reflection function $R(\tau;\mu,\mu_0,\phi-\phi_0)$ as  (e.g., Hansen and Travis, 1974) 
\begin{align}
I(\mu,\phi)&=R(\tau_{\rm T};\mu,\mu_0,\phi-\phi_0)\mu_0F_0\notag  \\
              & \hspace{1.5cm}\qquad(0\le \mu, \mu_0 \le 1) , \label{eq-2}  
 \end{align}
 where $F_0$ is the radiation flux  in units of  $\pi$ flowing per unit time through a unit area perpendicular to the direction of  incidence,  $\mu=\cos \theta$, and $\mu_0=\cos \theta_0$. The zenith angles $\theta$ and $\theta_0$ range from $0^\circ$ to $90^\circ$ and are measured with respect to the outward normal to the upper surface of the atmosphere $\Vec{n}$. (The transmission function $T(\tau_{\rm T};\mu,\mu_0,\phi-\phi_0)$ can be similarly  defined, but it is not restated here.) \par
To reduce the computational burden, let us expand  the reflection function as well as  other related quantities by using a Fourier series of  $\phi-\phi_0$, such that
\begin{align}
R(\tau; \mu, \mu_0, \phi-\phi_0) =&\displaystyle{\sum_{m=0}^M(2-\delta_{m0})R^m(\tau; \mu, \mu_0)\times} \notag \\
 &\qquad  \displaystyle{\times\cos m(\phi-\phi_0)}, \label{eq-3}
\end{align} 
where $\delta_{0m}$ designates the Kronecker delta. The Fourier coefficient of the reflection function $R^m(\tau; \mu, \mu_0)$ then satisfies the following invariant imbedding equation (Sato {\it et al.} 1977) :
\begin{align}
\displaystyle{{\partial R^m(\tau;\mu,\mu_0) \over \partial\tau} }&\displaystyle{ =-\left({1\over \mu}+{1\over \mu_0}\right)R^m(\tau;\mu,\mu_0)+} \notag \\
& \displaystyle{\hspace{1.5cm} +S^m(\tau; \mu, \mu_0),  } \label{eq-4}
\end{align}
where the source function $S^m(\tau; \mu, \mu_0)$ is defined as 
\begin{multline}
\hspace*{-0.3cm}\displaystyle{S^m(\tau;\mu,\mu_0)}=\displaystyle{ {1\over 4\mu\mu_0}P^m(\tau;-\mu,\mu_0)
 +{1\over 2\mu}\int_0^1P^m(\tau;  }\\
\mu,\mu')R^m(\tau;\mu',\mu_0){\rm d}\mu'  
 \displaystyle{+{1\over 2\mu_0}\int_0^1R^m(\tau;\mu,\mu')P^m(\tau; } \\
\mu',\mu_0){\rm d}\mu'
  \displaystyle{+\int_0^1R^m(\tau;\mu,\mu')\left[\int_0^1P^m(\tau;-\mu',\mu'')\times\right.}\displaybreak[0]\\
\hspace*{0.3cm}  \left.\phantom{\int_0^1}\times R^m(\tau; 
 \mu'',\mu_0){\rm d}\mu''\right]{\rm d}\mu',   \label{eq-5} 
 \end{multline} 
 together with the initial condition given at the ground surface by 
 \begin{equation}
 R^m(0;\mu,\mu_0)=A_{\rm grd}\delta_{m0}\quad (0\le A_{\rm grd} \le 1).  \label{eq-6}
 \end{equation}
 The functions $P^m(\tau;-\mu,\mu_0)$ and $P^m(\tau;\mu,\mu_0)$ in Eq.(\ref{eq-5})  are the Fourier coefficients of the mean phase function at the optical height $\tau$ which include the effect of the single scattering albedo of each scattering agent present therein :
 \begin{multline}
\displaystyle{P^m(\tau; u, u_0)=\sum_{k=0}^{K_{\rm s}}\varpi_{0,k}(\tau)\xi_k(\tau)P_k^m(\tau; u, u_0) } \\  (-1\le u, u_0\le 1).   \label{eq-7}
\end{multline}
\indent Note that  $k=0$ corresponds to atmospheric molecules causing Rayleigh scattering, whereas  $k\ge 1$ corresponds to  aerosols.  Also note  that we have employed the notations $u=\cos \vartheta$ and $u_0=\cos \vartheta_0$ defined with nadir angles $\vartheta$ and $\vartheta_0$, both ranging  from $0 ^\circ$ to $180^\circ$  
with respect to the downward normal $\Vec{n^\prime}$ (Fig. 1). The quantity $\varpi_{0,k}(\tau)$ is the single scattering albedo of  the $k$-th type scattering--absorbing agent located at a given optical height $\tau$, and  $\xi_k(\tau)$ is its fractional contribution  to the total extinction  coefficient per unit volume of the atmosphere there :
\begin{subequations}
\begin{align}
\varpi_{0,k}(\tau)&=\sigma_{{\rm s},k}(\tau)/\sigma_{{\rm e},k}(\tau)   \label{eq-8a}\\
\xi_k(\tau)&=\sigma_{{\rm e},k}(\tau),n_k(\tau)/\sum_{j=0}^{K_{\rm s}}\sigma_{{\rm e},j}(\tau)\,n_j(\tau)  
 \label{eq-8b}\\
\sigma_{{\rm e},k}(\tau)&=\sigma_{{\rm s},k}(\tau)+\sigma_{{\rm a},k}(\tau)
 \label{eq-8c}
\end{align} 
\end{subequations}
where  $\sigma_{{\rm s},k}(\tau)$,  $\sigma_{{\rm a},k}(\tau)$, and $\sigma_{{\rm e},k}(\tau)$ represent the scattering coefficient,  absorption coefficient, and extinction coefficient  per $k$-th type particle, respectively, and $n_k(\tau)$ represents the volume number density of the $k$-th type particles  at $\tau$.
The Fourier coefficient of the phase function for the $k$-th type  particles  $P^m_k(\tau; u, u_0) $ in Eq.(\ref{eq-7}) is given by
\begin{align}
P^m_k(\tau; u, u_0)&\displaystyle{={1\over \pi}\int_0^\pi\!\!\!P_k[\tau; \Theta(u, u_0, \phi-\phi^\prime)]\times} \notag \\
&\times\cos(\phi^\prime-\phi_0){\rm d}\phi^\prime,   \label{eq-9}
\end{align}
where $\Theta$ is the scattering angle specified by 
\begin{align}
\cos \Theta(u, u_0, \phi-\phi_0)& =uu_0+\sqrt{(1-u^2)(1-u_0^2)}\times \notag \\
& \qquad \qquad \times\cos(\phi-\phi_0).  \label{eq-10}
\end{align}
\indent Because  each of the $N$ slabs comprising the atmosphere is assumed  to be homogeneous,   these $\tau$-dependent quantities can be kept constant within a given slab.  \par
 To obtain the numerical solution of Eq.(\ref{eq-4}), we discretize it using the $N_\theta$-th order Gauss --Legendre quadrature points and their corresponding weights for
 performing the numerical integrations over $\mu^\prime$ and $\mu^{\prime\prime}$. The solutions for $R^m(\tau_{\rm T}; \mu, \mu_0)$ are then generated at the mesh points specified by the quadrature points on a square matrix of size unity. To these, we may also add
a certain number ($N_{\rm ext}$)  of non-quadrature $\mu$-points for the convenience 
of interpolations of  $R^m(\tau_{\rm T}; \mu, \mu_0)$ tables to obtain the emergent intensity of reflected light, $I(\mu, \phi)$, for a given angular set of  $(\mu, \mu_0, \phi-\phi_0)$.
\subsection{Fast Invariant Imbedding Method}
By introducing a new variable $t$ defined as
\begin{equation}
t=\tau-\tau_{{\rm tot},n-1}, \hspace{2cm} \label{eq-11}
\end{equation}
to specify the given optical height $\tau$ in terms of the height measured from the bottom of each slab, 
Eq.(\ref{eq-4}) can be  written in the short hand notation as
\begin{align}
\displaystyle{\frac{\partial R(t+\tau_{{\rm tot},n-1})}{\partial t}}&=-CR(t+\tau_{{\rm tot},n-1})+\notag \\
&\hspace{0.7cm} +S(t+\tau_{{\rm tot},n-1}),  \label{eq-12}
\end{align}
where the constant $C$ denotes $1/\mu+1/\mu_0$, and the index $n$ indicates the process of finding  the solution  for an  atmosphere  consisting of the first  $n$ slabs.  
Note that for large values of $C$, corresponding to a highly slanted incident or emergent 
radiation, this equation becomes {\it stiff}\ (see, e.g., Press {\it et al.} 1992 for a detailed discussion).
Therefore, obtaining the solution with sufficient accuracy requires  some  elaborate numerical tactics.
In the remaining sections, we shall delineate our method.\par
Given the value of  $R(t_j+\tau_{{\rm tot},n-1})$ at the $j$-th division point $t_j$ within the $n$-th slab, the solution at the next step $t_{j+1}$, i.e.,   $R(t_{j+1}+\tau_{{\rm tot},n-1})$, is given by
\begin{multline}
R(t_{j+1}+\tau_{{\rm tot},n-1})  \\
=\displaystyle{ R(t_j+\tau_{{\rm tot},n-1})
\exp[-C\cdot (t_{j+1}-t_j)]+ }\\
    \displaystyle{ +\int_{t_j}^{t_{j+1}}\!\!S(t+\tau_{{\rm tot},n-1})
\exp[{-C(t_{j+1}-t)}]{\rm d}t }  \label{eq-13} 
\end{multline}
with  $t_1=0$. This equation should be used recursively until the solution $R(\tau_{{\rm tot},n})$ is obtained.
The initial condition at $t_1$ is equal to the solution of an atmosphere having  $(n-1)$-slabs, i.e.,  $R(t_1+\tau_{{\rm tot},n-1})=R(\tau_{{\rm tot},n-1})$.\vspace{0.2cm}\\
(i) Solution at  $t_2\quad (j=1)$: \par 
The source function $S(t+\tau_{{\rm tot},n-1})$ is approximated by the first--order Lagrange polynomial of $t$  passing through two points $(t_1, S(t_1+\tau_{{\rm tot},n-1}) )$ and $(t_2, S(t_2+\tau_{{\rm tot},n-1}) )$ as  
\begin{multline}
S(t+\tau_{{\rm tot},n-1})\displaystyle{={1\over t_{21}}\left\{-S(t_1+\tau_{{\rm tot},n-1}) \times\right. }  \\
\displaystyle{\left.\times (t-t_2)+S(t_2+\tau_{{\rm tot},n-1}) (t-t_1)\right\} }  \label{eq-14}
\end{multline}
with $t_{21}\equiv t_2-t_1$. \par
Substituting Eq.(\ref{eq-14}) into Eq.(\ref{eq-13}) with $j=1$ and analytically integrating over $t$, we obtain  
 \begin{equation}
  R(t_2+\tau_{{\rm tot},n-1})=F_a+F_b\cdot S(t_2+\tau_{{\rm tot}, n-1})    \label{eq-15}
\end{equation}
where   
\begin{subequations}
\begin{align}
F_a&= R(t_1+\tau_{{\rm tot},n-1})E_{21}+\notag \\
&+{1\over C}\left(f_{21} -E_{21}\right) S(t_1+\tau_{{\rm tot},n-1}),  \label{eq-16a}\\
F_b&={1\over C}\left(1-{f_{21}} \right) , \label{eq-16b}\displaybreak[0]\\
f_{21}&= {1\over C \tau_{21}}\left(1-E_{21}\right),  \label{eq-16c}\\
E_{21}&=\exp(-C t_{21}). \label{eq-16d} 
\end{align}
\end{subequations}
\indent Although the value of $F_a$  is already known,  $S(t_2+\tau_{{\rm tot},n-1})$ involves the unknown $R(t_2+\tau_{{\rm tot},n-1})$. Hence,  Eq.(\ref{eq-15}) is a nonlinear implicit equation for $R(t_2+\tau_{{\rm tot},n-1})$, which is solved  by  successive iterations starting with  an obtained approximation, e.g., by setting $S(t_2+\tau_{{\rm tot},n-1})=S(t_1+\tau_{{\rm tot},n-1})$, i.e.,  
\begin{multline}
R(t_2+\tau_{{\rm tot},n-1})=R(t_1+\tau_{{\rm tot},n-1})\times\\
\times\exp(-C t_{21})+t_{21}f_{21}\cdot S(t_1+\tau_{{\rm tot},n-1}).  \label{eq-17}
\end{multline}
\indent The iterations are terminated when the condition 
\begin{equation}
\displaystyle{{\rm Max}. \left| 1- \frac{R^{\rm old}(t_2+\tau_{{\rm tot},n-1}) }{R^{\rm new}(t_2+\tau_{{\rm tot},n-1})}\right|\le \varepsilon_1   }  \label{eq-18}
\end{equation}
is   satisfied for  every combination  of  $\mu$- and $\mu_0$-quadrature points, where $\varepsilon_1$ designates the prescribed maximum relative error. \vspace{0.2cm}\\
(ii) Solution at  $t_3$\quad (j=2): \par
To increase the efficiency of integration over $t$, we take the step size that is larger than $t_{21}$ by a factor $\alpha_1\ (> 1)$, to obtain 
\begin{equation}
t_3=t_2+\alpha_1 t_{21}=(1+\alpha_1)\  t_{21}\quad (>2\,t_{21}).  \label{eq-19}
\end{equation}
\indent The source function $S(t+\tau_{{\rm tot},n-1})$ is approximated with the  quadratic Lagrange polynomial of $t$ that passes  through  the three points $(t_1, R(t_1+\tau_{{\rm tot},n-1}))$,  $(t_2, R(t_2+\tau_{{\rm tot},n-1}))$, and $(t_3, R(t_3+\tau_{{\rm tot},n-1}))$, viz., 
\begin{multline}
S(t+\tau_{{\rm tot},n-1})\\
\displaystyle{=S(t_1+\tau_{{\rm tot},n-1})\cdot(t-t_2)(t-t_3)/(t_{21}t_{31})- }\\
 \displaystyle{-S(t_2+\tau_{{\rm tot},n-1})\cdot(t-t_3)(t-t_1)/(t_{21}t_{32})+     }  \\
 \displaystyle{+S(t_3+\tau_{{\rm tot},n-1})\cdot(t-t_1)(t-t_2)/(t_{31}t_{32}),  }  \label{eq-20}
\end{multline}
where we have set $t_{21}=t_2-t_1$, $t_{31}=t_3-t_1$, and $t_{32}=t_3-t_2$.\par
Note that the values of $S(t_1+\tau_{{\rm tot},n-1})$ and $S(t_2+\tau_{{\rm tot},n-1})$ are now explicitly known.
Substitution of Eq.(\ref{eq-20}) into Eq.(\ref{eq-13}), again upon analytical integration over $t$,
 yields 
\begin{equation}
R(t_3+\tau_{{\rm tot},n-1})=F_a+F_b\cdot S(t_3+\tau_{{\rm tot},n-1}),  \label{eq-21}
\end{equation}
where  
\begin{subequations}
\begin{align}
F_a&=\displaystyle{ R(t_2+\tau_{{\rm tot},n-1})\exp(-C t_{32})+}\notag\\ 
&+H_1\cdot S(t_1+\tau_{{\rm tot},n-1})+\notag    \\
     & +H_2\cdot S(t_2+\tau_{{\rm tot},n-1}),    \label{eq-22a}\displaybreak[0]\\
F_b&=\displaystyle{(2f_{32}+E_{32}t_{21}-(t_{31}+t_{32})+}\notag \\
     & +Ct_{31}t_{32})/(C^2t_{31}t_{32}),\label{eq-22b)}\displaybreak[0]\\
H_1&= (2f_{32}-(1+E_{32})t_{32})/(C^2t_{21}t_{31}), \label{eq-22c}\displaybreak[0]\\
H_2&= (-2f_{32}+t_{31}+(t_{32}-t_{21}-\notag \\
     & -Ct_{21}t_{32})E_{32})/(C^2t_{21}t_{32}),  \label{eq-22d} \displaybreak[0] \\
f_{32}&=(1-E_{32})/C,  \label{eq-22e}  \\
E_{32}&=\exp(-Ct_{32}).   \label{eq-22f}       
\end{align}
\end{subequations}
\indent Because $S(t_3+\tau_{{\rm tot},n-1})$ is a function of  unknown $R(t_3+\tau_{{\rm tot},n-1})$, Eq.(\ref{eq-21}) is also a nonlinear implicit equation for $R(t_3+\tau_{{\rm tot},n-1})$. Therefore, we solve for $R(t_3+\tau_{{\rm tot},n-1})$ by successive iterations starting with an initial approximation given by a linear extrapolation of $R(t_1+\tau_{{\rm tot},n-1})$ and $R(t_2+\tau_{{\rm tot},n-1})$:
\begin{multline}
R(t_3+\tau_{{\rm tot},n-1})=\displaystyle{ \left\{(t_3-t_1) R(t_2+\tau_{{\rm tot},n-1})-\right.}\\
\displaystyle{\left.-(t_3-t_2) R(t_1+\tau_{{\rm tot},n-1})\right\}/t_{21} }  \label{eq-23}. 
\end{multline} 
\indent The iterations are terminated as soon as the condition 
\begin{equation}
\displaystyle{{\rm Max}.  \left|1-\frac{R^{\rm old}(t_3+\tau_{{\rm tot},n-1}) }{R^{\rm new}(t_3+\tau_{{\rm tot},n-1}))}\right|\le \varepsilon_1 } \label{eq-24}
\end{equation}
is  fulfilled similar to procedure (i).\vspace{0.2cm}\\
(iii) Solution at $t_4$ and above \ \ $(j\ge 3)$: \par
To obtain $R(t_4+\tau_{{\rm tot},n-1})$, for instance,  we  first relocate the foregoing solutions and related quantities  such that 
\begin{multline}
t_2, R(t_2+\tau_{{\rm tot},n-1}), S(t_2+\tau_{{\rm tot},n-1})   \\
              \Rightarrow \tau_1, R(t_1+\tau_{{\rm tot},n-1}) , S(t_1+\tau_{{\rm tot},n-1}), \nonumber 
\end{multline}   \vspace{-1cm}
\begin{multline}          
t_3, R(t_3+\tau_{{\rm tot},n-1}), S(t_3+\tau_{{\rm tot},n-1})  \\
              \Rightarrow t_2, R(t_2+\tau_{{\rm tot},n-1}) , S(t_2+\tau_{{\rm tot},n-1}). \nonumber
\end{multline}
\indent Then, we employ  $t_3+\alpha_1\,t_{32}$ as a new value for $t_3$
 and return to  procedure (ii). From this process, the new solution is always obtained  as $R(t_3+\tau_{{\rm tot},n-1})$.  The process is repeated until $t_3=\Delta\tau_n$ is attained. \par
If, however, the successive approximation for the solution at any step does  not converge within a prescribed number of iterations $N_{\rm iter}$,  the integration step size must be reduced by a certain factor $\alpha_2\quad (<1)$ before renewing the iteration.  \par
Furthermore, the integration of Eq.(\ref{eq-12}) over $t$ may be terminated whenever the maximum absolute value of the derivatives of $R(t+\tau_{{\rm tot},n-1})$ with respect to $t$ falls below a preset value $\varepsilon_2$:
\begin{align}
{\rm Max}. &\displaystyle{\left|{\partial R(t+\tau_{{\rm tot},n-1})/\partial t}\right|}\notag \\
&\displaystyle{={\rm Max}. \left|-CR(t+\tau_{{\rm tot},n-1}) +\right. } \notag \\
&\displaystyle{\left. +S(t+\tau_{{\rm tot},n-1})\right|\le \varepsilon_2}  \label{eq-25}
\end{align}
\subsection{Doubling--Adding  Method}
We  implement the doubling-adding  method to determine reflection function for the lowermost slab of the atmosphere of  interest to improve the computational efficiency. \par
Assume that the reflection and transmission functions $R^m(\tau; \mu, \mu_0)$ and $T^m(\tau; \mu, \mu_0)$  for a homogeneous  layer of optical thickness $\tau$ are known. Then we can obtain $R^m(2\tau; \mu, \mu_0)$ and $T^m(2\tau; \mu, \mu_0)$, viz., the reflection and transmission functions of a homogeneous layer 
 of the same optical properties but of optical thickness $2\tau$ by using equations  of the doubling method (e.g., Hansen and Travis, 1974): 
 \begin{subequations}
\begin{align}
R^m(2\tau; \mu, \mu_0)&=R^m(\tau; \mu, \mu_0)+ \notag \\
& +\exp(-\tau/\mu)U(\mu, \mu_0)+ \notag \\
& \displaystyle{+2\int_0^1T^m(\tau; \mu, \mu^\prime)U(\mu^\prime, \mu_0)\mu^\prime{\rm d}\mu^\prime}  \label{eq-26a}  \displaybreak[0]\\
 T^m(2\tau; \mu, \mu_0)&=\exp(-\tau/\mu)D(\mu, \mu_0)+\notag \\
&+T^m(\tau; \mu, \mu_0)\exp(-\tau/\mu_0)+ \notag \displaybreak[0]\\
 & \displaystyle{+2\int_0^1T^m(\tau; \mu, \mu^\prime)D(\mu^\prime, \mu_0)\mu^\prime{\rm d}\mu^\prime}  \label{eq-26b}
\end{align}
\end{subequations}
where
\begin{mathletters}
\begin{align}
Q_1(\mu, \mu_0)&=2\int_0^1R^m(\tau; \mu,\mu^\prime)R^m(\tau; \mu^\prime, \mu_0)\mu^\prime{\rm d}\mu^\prime, \label{eq-27a}  \\
Q_n(\mu, \mu_0)&=2\int_0^1Q_1(\mu,\mu^\prime)Q_{n-1}(\mu^\prime, \mu_0)\mu^\prime{\rm d}\mu^\prime \notag \\
&\hspace{3cm} (n\ge 2), \label{eq-27b} \displaybreak[0]\\
S(\mu, \mu_0)&=\displaystyle{\sum_{n=1}^\infty Q_n(\mu,\mu_0),} \label{eq-27c}\displaybreak[0]\\
D(\mu, \mu_0)&=\displaystyle{T^m(\tau; \mu,\mu_0)+S(\mu,\mu_0)\exp(-\tau/\mu_0)+} \notag \\
 & \displaystyle{+2\int_0^1S(\mu,\mu^\prime)T^m(\tau;\mu^\prime,\mu_0)\mu^\prime{\rm d}\mu^\prime,} \label{eq-27d} \displaybreak[0]\\
 U(\mu, \mu_0)&= \displaystyle{R^m(\tau; \mu,\mu_0)\exp(-\tau/\mu_0)+} \notag \\
 & \displaystyle{+2\int_0^1R^m(\tau; \mu,\mu^\prime)D(\mu^\prime,\mu_0)\mu^\prime{\rm d}\mu^\prime.} \label{eq-27e}
 \end{align}
 \end{mathletters}
\indent We repeat the above procedure until the desired value of $\Delta\tau_1$ is reached.  Note that
 the reflection and transmission functions thereby obtained do not consider the effect of  ground reflectivity.  If the atmosphere is bounded at its bottom by a Lambert surface of 
reflectivity $A_{\rm grd}$, its effect manifests through the azimuth angle-independent  Fourier coefficient $R^0(\Delta\tau_1; \mu, \mu_0)$, and the adding method gives rise to  the following  expression\,(p.64 of van de Hust, 1980) :
\begin{align}
R_{\rm S}^0(\Delta\tau_1; \mu, \mu_0)&=R^0(\Delta\tau_1; \mu,\mu_0)+\notag \\
&\displaystyle{+{A_{\rm grd}\over 1-A_{\rm grd}A_{\rm sph}}t_a(\mu)t_a(\mu_0) }  \label{eq-28}
\end{align}
where  $A_{\rm sph}$ is the spherical or Bond albedo defined as
\begin{equation}
\displaystyle{A_{\rm sph}=4\int_0^1\!\!\!\int_0^1\!R^0(\Delta\tau_1; \mu, \mu_0)\mu\mu_0{\rm d}\mu{\rm d}\mu_0,}  \label{eq-29}
\end{equation}
and the function $t_a(\mu)$ is given  by
\begin{equation}
\displaystyle{t_a(\mu)=\exp(-\Delta\tau_1/\mu)+2\int_0^1T^0(\Delta\tau_1; \mu, \mu^\prime)\mu^\prime{\rm d}\mu^\prime.  }  \label{eq-30}
\end{equation}
To initialize the doubling calculation, we start with a slab of  optical thickness $\tau_{\rm in}$ given by
\begin{equation}
\tau_{\rm in}=\Delta\tau_1/2^{N_0+N_{\rm D}},   \label{eq-31}
\end{equation} 
where  $N_0={\rm int}\{\log_{10}\Delta\tau_1/\log_{10}2\}$\footnote{The symbol {\rm int}\,$\{x\}$ used here  signifies Gauss' symbol, i.e., the greatest integer that is equal to or less than $x$.}, and $N_{\rm D}$ is a prescribed integer.   \par
The reflection and transmission functions for this thickness are assumed to be sufficiently well approximated by the sum of the single and  second--order scattering solutions (e.g., Kawabata and Ueno, 1988).
\section{Multiple Scattering  Calculations with Current Method}
\subsection{Setting up Numerical Calculations }
Following Sato {\it et al.}\,(1977), we use  the Venus cloud model of Hansen and Hovenier (1974) at a  wavelength 365 nm. This model was derived by them on the basis of a theoretical analysis of  ground-based polarimetry data.
Briefly, the cloud is a thick layer consisting of homogeneous mixture of  ${\rm CO_2}$ molecules and droplets of concentrated aqueous sulfuric acid that have a spherical shape. The real part of  refractive index  $n_r$  of these droplets at this wavelength is assumed to be 1.46, and its imaginary part $n_i$ is assumed to be 0.   
The size distribution of the  radius $r$ of the cloud particleis is approximated by the gamma distribution characterized by an effective radius $r_{\rm eff}$ of 1.05 $\mu$m and an effective variance $v_{\rm eff}$ of 0.07:
\begin{equation}
\displaystyle{n(r)={(ab)^{(2b-1)/b}\over \Gamma[(1-2b)/b] } r^{(1-3b)/b}\exp(-r/ab) , }  \label{eq-32}
\end{equation}
where $a=r_{\rm eff}$, $b=v_{\rm eff}$, and $\Gamma$ is the gamma function (Hansen and Travis, 1974). \par
The phase function for the cloud particles averaged over this size distribution can be 
generated by a Mie scattering computer code. \par 
However, the UV absorbers that are definitely present in the actual clouds are completely ignored in this study to maximize the effect of multiple scattering of light. Therefore, the single scattering albedos $\varpi_{0,k} \ (k=0, 1)$, i.e., Eq.(\ref{eq-8a}), of the  molecules and aerosol particles are set to be unity. The extinction fraction $\xi_0$ due to  Rayleigh scattering  by ${\rm CO_2}$ molecules is assumed to be 0.04.
Thus, the extinction  fraction by sulfuric acid cloud particles $\xi_1$ is 0.96\,(Eq.(\ref{eq-8b})). \par 
The optical thickness of the cloud layer is assumed to be 128 in the current study, and a  Lambert surface with 
$A_{\rm grd}=1$ is placed at its bottom. \par
The Fourier sum indicated in Eq.(\ref{eq-3}) for $R(\tau; \mu, $\\ 
$\mu_0, \phi-\phi_0)$ is terminated at  $M=34$, and a 150-point Gauss--Legendre quadrature is applied to integrate Eq.(\ref{eq-9}) over $\phi^\prime$ to obtain the Fourier coefficients of the phase function. \vspace{0.2cm}\\
A)  Fast Invariant Imbedding Calculations \par
On the basis of  various past experiments,  we adopt the following values for the relevant parameters:\\
\hspace*{0.5cm}\begin{tabular}{lllll}
$N_\theta=29$, &$N_{\rm iter}=30$, &  $\tau_{21}=10^{-2}$, &      \\
$\alpha_1=1.2$, & $\alpha_2=0.8$, &      &        \\
$\varepsilon_1=10^{-8}$, & $\varepsilon_2=10^{-10}$. &         & 
\end{tabular}\vspace{0.2cm}\\ 
\indent To choose a suitable value for $N_\theta$ (the order of  the Gauss--Legendre quadrature for $\mu$-integrations), we varied the value of $N_\theta$ in some sample calculations of the intensity distribution  $I(\mu, \phi)/F_0$ for the  reflected sunlight along the intensity equator of a spherical planet viewed from an infinite distance with a phase angle of $5^\circ$. 
For this purpose, we first defined a Cartesian coordinate system on a projected planetary disk of unit radius such that the $x$-axis ran along the intensity equator and the $y$-axis  ran perpendicular to it at the disk center which corresponds to the sub-observer point. \par 
The scattering geometry at a given location $(x, y)$ on the disk
 can  then be specified by the following equations\, (e.g., Kawabata {\it et al.} 2000):
\begin{subequations}
\begin{align}
\mu&=\sqrt{1-(x^2+y^2)},   \label{eq-33a} \\
\mu_0&= \mu\cos \alpha +x \sin \alpha, \label{eq-33b} \displaybreak[0]\\
\cos (\phi-\phi_0)&= (\mu\mu_0-\cos \alpha)/A, \label{eq-33c}\displaybreak[0] \\
\sin (\phi-\phi_0)&= y\sin \alpha/A,  \label{eq-33d}\\
A&=\sqrt{(1-\mu^2)(1-\mu_0^2)}, 
\end{align}
\end{subequations}
where $\alpha$ is the phase angle.   \par
The intensity equator corresponds to $y=0$, and the sub-solar point is located at $(\sin \alpha, 0)$. The  bright limb and the terminator of the planetary disk intersect with the $x$-axis at $(1,0)$ and $(-\cos \alpha,0)$, respectively, for positive values of $\alpha$.  \par
For the geometry $(\mu, \mu_0, \phi-\phi_0)$ associated with a given location  $(x, 0)$, the square tables of the Fourier coefficient of the reflection function  $R^m(\tau_{\rm T};\mu, \mu_0)$ are interpolated at  $(\mu, \mu_0)$ by using the bicubic interpolation method (Press {\it et al.} 1992).  Then the results are  summed up according to Eq.(\ref{eq-3}) to produce $R(\tau_{\rm T}; \mu, \mu_0, \phi-\phi_0)\mu_0$, which is just equal to the emergent intensity $I(\mu, \phi)/F_0$ of the reflected sunlight at the point in question. \\
Furthermore, we add two extra $\mu$-points, viz., 0.1 and 1, to conveniently interpolate 
 the $R^m$ tables.\par
\begin{figure}[!b]
\centering
\vspace*{1cm}\includegraphics[width=0.98\linewidth]{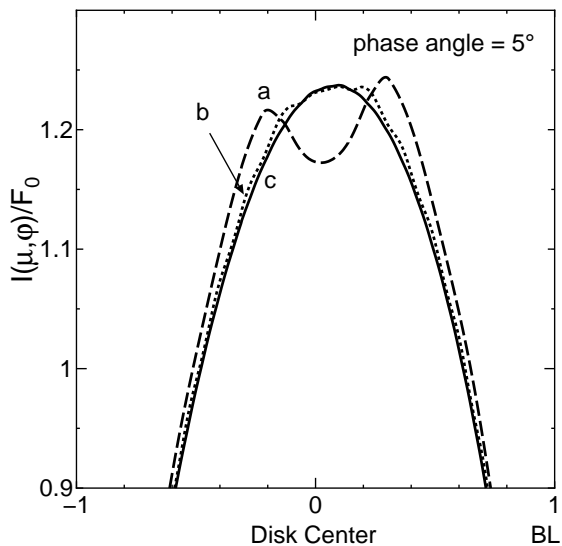}
\vspace*{0.5cm}
\caption{Calculated emergent intensity distributions  along the intensity equator of a planetary disk.
The curves a, b, and c are the theoretical  distributions of  the intensity $I(\mu, \phi)/F_0$ along the intensity equator of a spherical planet for
 a phase angle of $5^\circ$; they were  calculated adopting  a Gauss--Legendre quadrature of $N_\theta=7$, $15$, and $29$, respectively. Note that  $F_0$ is the flux of the incident radiation in units of  $\pi$\,(only the portions with $I(\mu, \phi)/F_0 \ge 0.9$ are shown). The locations $x=0$ and 1, respectively,  correspond to the disk center\,(or sub-observer point) and the bright limb\,(BL). A Hansen--Hovenier Venus model cloud  for a wavelength of 365\,nm was employed, except that the single scattering albedo was set to unity, and the bottom was bounded by a perfectly reflecting Lambert surface, as  described in the text. 
 \label{fig2}}
\end{figure}
Fig. 2 shows three intensity distributions $I(\mu, \phi)/F_0$ calculated  along the intensity equator for  $N_\theta =7$, $15$, and $29$ (only the portions with $I(\mu, \phi)/F_0 \ge 0.9$ are displayed).  
The model atmosphere employed for these calculations is the Hansen--Hovenier Venus cloud
model consisting of ${\rm CO_2}$ molecules and aerosol particles of concentrated sulfuric acid
as described previously in this section. \par
Obviously, it is imperative that $N_\theta$ should be sufficiently large to obtain reliable theoretical intensity distributions.  For this reason, we adopt $N_\theta=29$ for subsequent multiple scattering calculations. \par
Note that obtaining the numerical solution of Eq.(\ref{eq-12}) by means of, e.g., the ordinary  fourth-- order Runge--Kutta method requires a step size $t_{21}$ comparable to or much smaller than the minimal value of  $C^{-1}$. With $N_\theta=29$, we would therefore have to employ a value less than or equal to  $10^{-3}$ for $t_{21}$. \par 
Furthermore,  we must  maintain  $\alpha_1=1$ for the step size throughout the process of integration until the entire atmosphere is completed. In fact, the CPU time required for the Runge--Kutta method to obtain  $R^0(128; \mu, \mu_0)$ for the Hansen--Hovenier Venus model cloud of optical thickness 128 but with the unit single scattering albedo is found to be larger than that required by the fast invariant imbedding method by more than a factor of  900. \par
Therefore, it is impractical to employ the Runge--Kutta method as a numerical solver of the invariant imbedding equations.       
\newline\\
B) Doubling--Adding Calculations\par
First of all, our current method requires the doubling--adding method to produce the reflection function for the lowermost slab of optical thickness $\Delta\tau_1$, although the use of the adding method is rather implicit in this case due to the fact that a Lambert plane is assumed for the bottom surface.
As in A), we employ $N_\theta=29$ for the Gauss--Legendre quadrature to perform the $\mu$-integrations involved in Eqs.(26) and (27). \par 
To set up the value for $\tau_{\rm in}$, we adopt $N_{\rm D}=25$ for Eq.(\ref{eq-31}), which implies that the starting solutions for the reflection and transmission functions $R^m(\tau_{\rm in}; \mu, \mu_0)$ and $T^m(\tau_{\rm in}; \mu, \mu_0)$, respectively, are generated for a homogeneous layer of an optical thickness of  the order of  $10^{-8}$, by summing  the single and second--order scattering solutions  using the expressions of Kawabata and Ueno (1988).   For the  Fourier summation of Eq. (\ref{eq-3}),   $M=34$ is adopted as is done for the fast invariant imbedding method. \par
Secondly, for comparison, we also perform the doubling--adding calculations with the same parameter values as indicated above employing the model atmospheres used to test the current method.
\subsection{Numerical Comparison}
The filled circles in Fig. 3 show the ratio of the CPU time required by the fast invariant imbedding method  $t({\rm FII})$ to that required by the doubling--adding method $t({\rm DA})$ to obtain  
the first 35 Fourier coefficients $R^m(128; \mu, \mu_0)\quad (m=0, 1, 2, \cdots, 34)$ for a conservatively scattering Hansen--Hovenier cloud. This ratio is presented as a function of the degree of the Gauss--Legendre quadrature $N_\theta$. The cloud has an optical thickness of 128, and it is bounded by a perfect Lambert surface with $A_{\rm grd}=1$ at its bottom. The solid  curve is a cubic-polynimial least square fit to the data.  
\begin{figure}[!tb]
\centering
\includegraphics[width=0.98\linewidth]{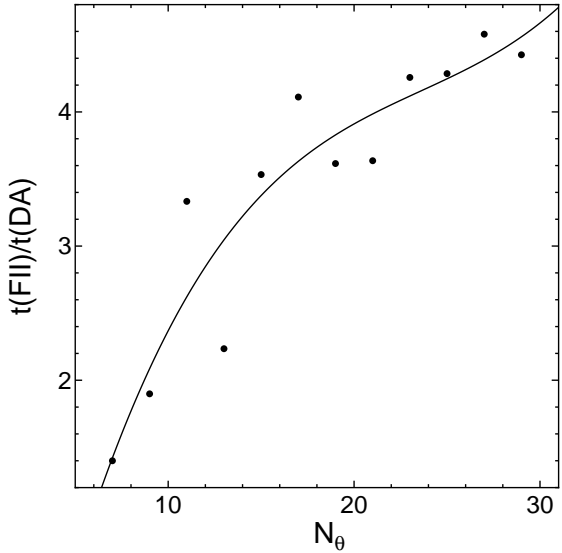}
\vspace*{0.5cm}
\caption{The CPU time $t({\rm FII})$ required for the fast invariant imbedding method to produce 
the reflection function $R(\tau_{\rm T}; \mu, \mu_0, \phi-\phi_0)$ of the model atmosphere described in $\S 3.1$
compared to the doubling--adding  method CPU time $t({\rm DA})$ as a function of the degree of the Gauss--Legendre quadrature, $N_\theta$.
  \label{fig3}}
\end{figure}
In  case of a single thick homogeneous atmosphere, the CPU time of the fast invariant imbedding method rapidly increases with the value of $N_\theta$: for  $N_\theta=29$,  it is  approximately five times slower than the doubling method.  In other words, the fast invariant imbedding method can hardly compete with the doubling method for a single thick homogeneous atmosphere at high orders of the Gauss--Legendre quadrature.\par
Fig.4 shows, on the plane of the optical thickness of each slab $\tau_{\rm T}/N$ versus the number of slabs $N$, a demarcation line along which the current hybrid method and the doubling--adding method work equally fast. 
Above this line (in the shaded area), the current hybrid method is faster than the doubling--adding method; below this line, the opposite is true.
 For a given number of  slabs,  all the slabs are set to equal optical thicknesses and
 optical properties identical to that of the Hansen--Hovenier Venus model cloud described in $\S 3.1$.
\begin{figure}[!tb]
\centering
\includegraphics[width=0.98\linewidth]{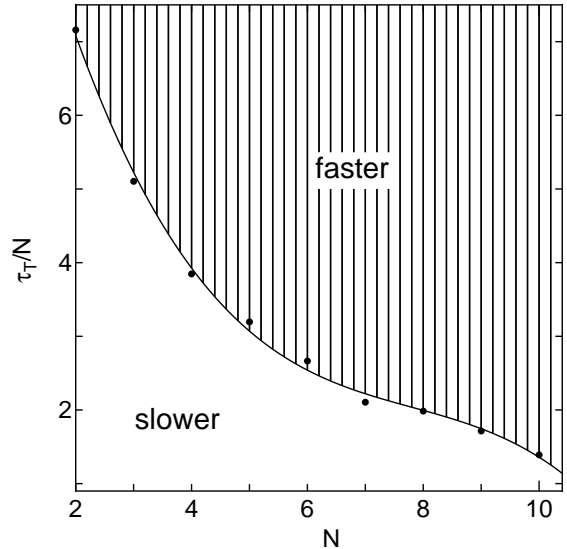}
\vspace*{0.5cm}
\caption{Demarcation line along which the current method works as fast as the doubling--adding method for $\tau_{\rm T}/N$ (the optical thickness of each slab)  versus $N$ (the number of identical slabs comprising the atmosphere).
The shaded area indicates the region where the hybrid method is faster than the doubling--adding method.
The optical properties of each slab are those employed for Fig.3. 
 \label{fig4}}
\end{figure}
The solid curve is a cubic--polynomial least  square fit to the data.  As the number of slabs increases, 
 the efficiency of the current method increases, and the shaded area extends to 
an increasingly smaller optical thickness associated with each slab.  This indicates that stacking up a large number of slabs by means of the current method is more rapid than by the doubling--adding method. \par
 \begin{table}[htb]
\caption{Reflection function calculated by the current and doubling--adding methods}
\begin{center}
\begin{tabular}{l|cc}
\hline\hline
     $R(\tau_{\rm T}; \mu, \mu_0,\phi-\phi_0)\mu_0$     &  current   &  doub--add.  \\  \hline
$R(35; 0.1, 0.1, 0^\circ)\times 0.1$   & 2.126698 & 2.126698  \\
$R(35;0.1, 0.1,180^\circ)\times 0.1$  & 0.246565 & 0.246562   \\
$R(35; 0.5, 0.5, 0^\circ)\times 0.5$   & 0.649196 & 0.649197  \\
$R(35; 0.5,0.5, 180^\circ)\times 0.5$ & 0.609809 & 0.609809  \\
$R(35; 1.0, 1.0, 0^\circ) $               &  1.258023 & 1.257902 \\ \hline   
\end{tabular}
\end{center}
\end{table}
\indent Fig. 5 shows the maximum value of  the CPU time of the current method relative to the doubling--adding CPU time (left-hand side ordinate) as a function of the number of homogeneous slabs comprising an atmosphere whose optical properties are the same as those employed for Fig. 4. The filled circles are the data points, and the solid curve is a B-spline fit.  The open circles are the data points for the right-hand side ordinate, which  indicates the optical thickness of each slab that yields the maximum CPU time, and the dashed curve is a B-spline fit to them.The optical properties of each slab are the same as those employed for  Fig. 4.  \par
\begin{figure}[!htb]
\centering
\includegraphics[width=0.98\linewidth]{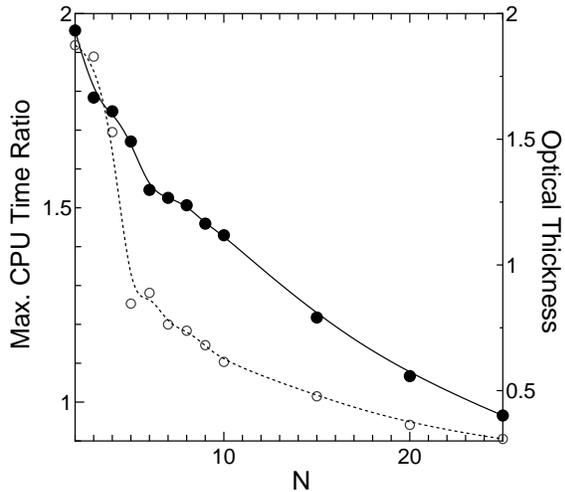}
\vspace*{0.5cm}
\caption{Maximum value of  the CPU time required for the current  method relative to the doubling--adding CPU time.
The abscissa indicates the number of slabs of equal thickness comprising the atmosphere employed, and the left-hand side ordinate shows  the maximum CPU time relative to the doubling--adding CPU time to perform the same set of multiple scattering calculations. The filled circles are the data points for the maximum CPU time ratio, and the solid curve is a B-spline fit to this data.  The open circles  are the data points for the optical thickness of each slab (refer to the right-hand side ordinate) that gives the maximum CPU time, and the dashed curve is a B-spline fit to this data.
The optical properties of each slab are the same as those employed for  Fig. 4. 
 \label{fig5}}
\end{figure}
The column  ''current'' in Table 1 shows  a set of sample values of $R(35; \mu_0, \mu_0, \phi-\phi_0)\mu_0$ obtained by the current hybrid method for five combinations of $\mu_0(=0.1, 0.5, {\rm  and}\,  1)$, and $\phi-\phi_0(=0^\circ {\rm and}\, 180^\circ)$. The model atmosphere employed is composed of  seven identical slabs, each having an optical thickness of 5 and the same optical properties as  those assumed for Fig. 4.   The column ''doub--add'' shows  the corresponding values of the reflection function produced by the doubling--adding method.  Note that even the largest discrepancy found for $\mu_0=1$ is  less than $10^{-2}$\,\%. \par 
\begin{figure}[!htb]
\includegraphics[width=0.88\linewidth]{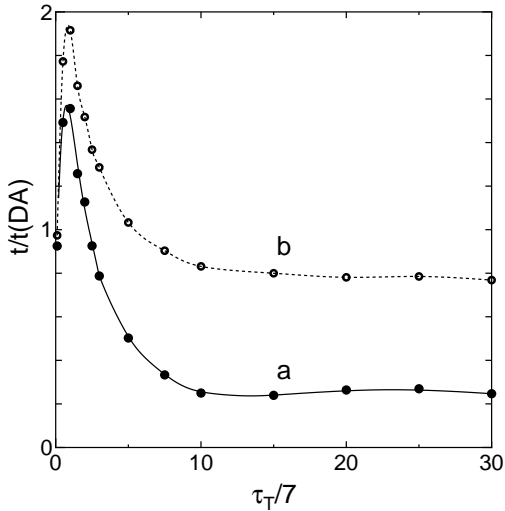}
\vspace*{0.4cm}
\caption{The CPU time of  (a) the current hybrid method and (b) the fast invariant imbedding method compared to the doubling-adding method CPU time $t({\rm DA})$ for an atmosphere consisting of seven identical slabs as a function of the optical thickness  $\tau_{\rm T}/7$ of each slab.
 The current method is found to be significantly more efficient than the doubling--adding method for $\tau_{\rm T}/7 \gtrsim 2.5$.  In addition, note that the execution speed of the hybrid method is greater  by almost a factor of four than  the doubling--adding  method for $\tau_{\rm T}/7 \gtrsim 10$.
 \label{fig6}}
\end{figure}
Fig. 6 shows the CPU time $t$ of (a) the current hybrid method and (b) the fast invariant imbedding method compared to that of the doubling--adding method $t({\rm DA})$
for an atmosphere composed of  seven identical slabs whose optical properties are the same as those for Fig. 4. The abscissa is the optical thickness of each slab $\tau_{\rm T}/7\ (=\Delta \tau_n,\ n=1, 2, \cdots, 7)$. The filled circles are the points for the current method, and the solid curve designated  by the letter {\it a} is a B-spline curve fit to them.
The open circles are the data points for the fast invariant imbedding method, and the dashed curve designated by the letter {\it b} is a B-spline fit to them.  \par
 For  $\tau_{\rm T}/7\gtrsim 2.5$, the hybrid method is definitely faster than the doubling--adding method and as the value of $\tau_{\rm T}/7$ increases, the hybrid method's CPU time  asymptotically approaches  approximately a quarter of that for the doubling--adding method.
 Although  the opposite is the case for $\tau_{\rm T}/7\lesssim 2.5$, the relative CPU time of the  current  method is not greater than  1.6 occurring at $\tau_{\rm T}/7 \simeq 0.7$.    \par
In contrast, the relative CPU time of the fast invariant imbedding method is lower than  that of the doubling--adding method only for $\tau_{\rm T}/7\gtrsim 6$ and approaches a limiting value of approximately  0.75 as $\tau_{\rm T}/7$ increases. This limiting value is, however,  almost a factor of three larger than  that of the hybrid method.  For optical thicknesses less than 6, the fast invariant imbedding method is slower than the doubling--adding method, and the relative CPU  time is 1.92 at $\tau_{\rm T}/7=1$ (as opposed to 1.6 at $\tau_{\rm T}/7=0.7$ for the current method as stated above).
These facts firmly  attest to the high practicability of the current  method as a computational tool for  remote sensing data analyses.      \par
Note that the results described above are for a stack of slabs of equal optical thickness. In actual model calculations, however, the lowermost slab is likely to  have the largest optical thickness. Therefore,
the efficiency of the current method in actual model calculations is higher than that observed in this section.
\section{Conclusion}
We have succeeded in creating a new and highly efficient method for multiple scattering calculations by coupling the fast invariant imbedding method with the doubling--adding method.  \par
Our new hybrid method  enhances the advantage of these two methods, while complementing their shortcomings. The fast
invariant imbedding method is for atmospheres composed of  a large number of slabs, but tends to be significantly slower for  atmospheres comprising a small number of relatively thick slabs.
In contrast, the speed of the doubling--adding method is slow for atmospheres composed of a large number of slabs, because the number of the  time-consuming adding calculations increases.  \par
   The execution speed of the new method  may still turn out to be  slower than the doubling--adding method, probably the fastest method proposed so far,
 in handling atmospheres stratified with  a relatively small number of  homogeneous slabs. For example, 
for  a two-slab atmosphere,  this hybrid method is slower than the doubling--adding method for the optical thicknesses less than 7 as observed from Fig.4. Even so, the CPU time  required is not more than 
 twice that required by the doubling--adding method. \par
Furthermore, for a larger number of slabs, the differences are likely to be much less significant, as shown in Figs. 5 and  6.   In fact, for $N\gtrsim 25$, the speed of the current method surpasses that of the doubling--adding method.  In addition, for a given number of slabs, this hybrid method is capable of working approximately four times faster than the doubling--adding method if the optical thickness of each layer is larger than a certain threshold  value,  as can be observed from Figs. 4 and 6. \par
All comparisons in this study are based on a stratified atmosphere consisting of slabs of equal optical thickness.  However, in actual models, the lowermost slab tends to have the largest optical thickness.
Under such circumstances, the hybrid method proposed in this study  should prove more advantageous than the doubling--adding method in performing multiple scattering calculations.  \\
\acknowledgments
\noindent{\bf Acknowledgment}\indent The author is grateful to the late Prof. Sueo Ueno for having directed his attention to the theory of radiative transfer. \par
This work has been published in {\it NAIS Journal}~{\bf 7} (ISSN 1882-9392), 5--16 (2012).
 \vspace{1.5cm}\\

\end{document}